\documentclass[final,11pt,times,numeric]{elsarticle}

\usepackage[tableposition=top]{caption}
\usepackage[usenames,dvipsnames]{color}
\usepackage{float}    \floatstyle{plaintop} \restylefloat{table} % Make captions on top of a table (where it should be)
\usepackage{physics}

\usepackage[title]{appendix}
\usepackage{amsmath}  % needed for \tfrac, \bmatrix, etc.
\usepackage{amsfonts} % needed for bold Greek, Fraktur, and blackboard bold
\usepackage{amssymb}
\usepackage{lipsum}
\usepackage{amsthm}
\usepackage{graphicx} % needed for figures
\usepackage{epsfig}
\usepackage{accents}
\usepackage{tikz} 
\usetikzlibrary{matrix,decorations.pathreplacing}
\usepackage{pgfplots}
\pgfplotsset{compat=1.18}

\definecolor{ao}{rgb}{0.0, 0.0, 1.0}
\definecolor{armygreen}{rgb}{0.29, 0.33, 0.13}

\newcommand{\sub}[1]{_{\stackrel{}{#1}}}
\newcommand{\lpp}{\stackrel{\leftarrow}{\partial}_{p}}
\newcommand{\rpp}{\stackrel{\rightarrow}{\partial}_{p}}
\newcommand{\lpx}{\stackrel{\leftarrow}{\partial}_{x}}
\newcommand{\rpx}{\stackrel{\rightarrow}{\partial}_{x}}
\newcommand{\lbb}{ \{{\hskip-2pt}\{ }
\newcommand{\rbb}{ \}{\hskip-2pt}\} }
\newcommand{\Pl}[1]{\partial_{\stackrel{}{#1}}}
\newcommand{\Grad}{{\vec\nabla}}

\newcommand{\cB}{{\cal B}}
\newcommand{\cD}{{\cal D}}

\newcommand{\cM}{{\cal M}}
\newcommand{\cP}{{\cal P}}
\newcommand{\cI}{{\cal I}}
\newcommand{\del}{{\partial}}

\newcommand{\pl}{{\partial}}
\newcommand{\Sr}{{Schr\"odinger}}

\journal{Annals of Physics}

\begin{document}

\begin{frontmatter}

\title{Form-preserving transformations of wave and Wigner functions}

\author{Mustafa Amin, Mason Daub, Mark A. Walton} 

\affiliation{
	organization={Department of Physics and Astronomy, University of Lethbridge},
	addressline={4401 University Drive West, Lethbridge, Alberta}, country={Canada\ \ T1K~3M4} 
 }

\begin{abstract}
Solutions of the time-dependent Schrödinger equation are mapped to other solutions for a (possibly) different potential by so-called form-preserving transformations. These time-dependent transformations of the space and time coordinates can produce remarkable solutions with surprising properties. A classic example is the force-free accelerating Airy beam found by Berry and Balazs. We review the 1-dimensional form-preserving transformations and show that they also yield Senitzky coherent excited states and the free dispersion of any harmonic-oscillator stationary state. Form preservation of the $D$- and 3-dimensional Schrödinger equation with both a scalar and a vector potential is then considered. Time-dependent rotations may be included when a vector potential is present; we find a general transformation formula for this case. Quantum form-preserving maps are also considered in phase space. First, the wave-function transformation is shown to produce a simple result for Wigner functions: they transform as a true phase-space probability would. The explicit transformation formula explains and generalizes the rigid evolution of curves in phase space that characterize the Airy beam and the coherent excited states. Then we recover the known form-preserving transformations from the Moyal equation obeyed by Wigner functions.
\end{abstract}


\end{frontmatter}

\section{Introduction}
\label{introduction}

The 1-dimensional time-dependent Schrödinger equation has some remarkable solutions.  A classic is the Airy beam found by Berry and Balazs \cite{BB79}. Even though the potential vanishes, it has features that move with a constant acceleration. 

Also striking are the coherent excited states of the harmonic oscillator found by Senitzky \cite{Se54}. Their wave functions have the instantaneous modulus of a stationary energy eigenstate. As time passes, that waveform oscillates rigidly and harmonically about the equilibrium point with arbitrary amplitude and phase. For the lowest energy, the waveform is Gaussian and the Senitzky wave function describes the famous coherent state. The coherent excited states are less well known \cite{Ph14}. 

The Airy beam and the coherent excited states describe very different physics but have clear formal similarities. Both can be produced from less exotic wave functions by what have been called form-preserving transformations. Preservation of form means that the time-dependent Schrödinger equation is mapped to another time-dependent Schrödinger equation. However, the original and transformed potentials may be different.  The space and time coordinates $(x,t)$ are transformed to $({x'}, {t'})$ so that a Schrödinger solution $\psi$ for a potential $V$ is mapped to another solution ${\psi'}$ for a potential ${V'}$.  

If $V'$ and $V$ coincide, the transformation is a symmetry of the \Sr\ equation. The symmetries of differential equations such as the Schrödinger equation are found by methods going back to Sophus Lie (for more contemporary treatments, see \cite{Olver00, Miller84, BlumanCole74}, for examples). The symmetry groups of Schrödinger equations for different potentials have been worked out this way. The free particle was first treated in \cite{Ni72}. For other nonvanishing potentials, see \cite{BSW76} and references therein. A historical treatment of \Sr\  symmetry is given in \cite{Duval2024}. 

The Lie methods are adapted straightforwardly to form-preserving transformations that intertwine solutions for different potentials. It was Niederer who found the first form-preserving transformations of the Schrödinger equations \cite{Ni73}.\footnote{Later, Takagi \cite{Ta90} independently found the same result, calling it the ``equivalence of a harmonic oscillator to a free particle''.}  He mapped the wave functions of the one-dimensional free particle into those of the harmonic oscillator after showing that the corresponding Schrödinger symmetry groups were isomorphic. This first example can be generalized to include an additional non-trivial potential, indicating that it is the transformation that is the important property. The most general form-preserving transformations of the one-dimensional \Sr\  equation were written explicitly in \cite{Fi99, BS96}. 

The next section examines remarkable \Sr\ wave functions obtained from the form-preserving transformations. In particular, 3 examples are discussed. To start, both the Senitzky excited coherent states \cite{Se54, Ph14} and the Berry-Balazs accelerating Airy beam \cite{BB79} are briefly discussed. The general \Sr\  form-preserving transformations are then written and we demonstrate that they produce the Berry-Balazs and Senitzky wave functions from well-known stationary states. In addition, we show that such a transformation derives the free dispersion of harmonic-oscillator stationary waveforms, using the map between the free particle and the harmonic oscillator. 

Section 3 considers Schrödinger form-preservation in multiple spatial dimensions \cite{GRo72, TaI91, Nikitin2020}. We find the general transformations retaining the form of the simplest $D$- and 3-dimensional Schrödinger equation that involves both a scalar and a vector potential. When a vector potential is present, the transformations can involve time-dependent rotations. 

Phase-space considerations are reported in Section 4.  First, the effect of the form-preserving transformation of \Sr\ wave functions on the corresponding Wigner functions is studied. A notably simple transformation formula for the Wigner functions is found.  We conjecture that it is the general rule for form-preserving transformations of Wigner functions. It explains and generalizes the rigid translation of phase-space curves associated with Airy beam wave function \cite{BB79} and the Senitzky wave functions. 

Also in Section 4, we analyze form-preserving transformations of the Moyal equation (see \cite{CFZ14, Ha04, ZFC05}, e.g.), the equation of motion of the Wigner function. Since phase-space quantum mechanics is an independent, stand-alone formulation of quantum mechanics, one should be able to complete such an analysis without referring to wave functions. We analyze the Moyal equation carefully, indicating why a general treatment is difficult. Restricting the possibilities somewhat, however, we were able to re-derive the general transformations that preserve the form of \Sr\ equations. 

Section 5 is our conclusion. It summarizes the work reported here briefly and mentions possible connections to current research.

\section{Remarkable time-dependent wave functions generated by form-preserving transformations} 

Closed-form exact solutions of the time-dependent \Sr\ equation are harder to come by than those of the time-independent \Sr\ equation.  This paper discusses a method of generating such solutions.   

We first focus on a pair of striking wave functions.

\subsection{Berry-Balazs and Senitzky solutions}

The Airy beam \cite{BB79} and the Senitzky coherent excited states \cite{Se54} are noteworthy solutions of the time-dependent    
Schrödinger equation 
\begin{align}
	i\hbar\frac{\partial\psi}{\partial t} = -\frac{\hbar^2}{2m}\frac{\partial^2\psi}{\partial x^2} + V\,\psi \ . 
\label{SE(t)}\end{align}
Here  $\psi(x, t)$ is the wave function and the potential energy is $V=V(x,t)$.  

The Airy beam has wave function 
\begin{align}
	\psi(x,t)\ =\  \text{Ai}\left[ \frac B{\hbar^{2/3} } \left(x - \frac{B^3t^2}{4m^2}\right)\right]\, \exp\left[ i\frac{B^3t}{2m\hbar}\left( x - \frac{B^3t^2}{6m^2}\right)  \right]\  .
\label{BBwf}\end{align}
Although the argument of the Airy function shows that the wave function has features that accelerate, it satisfies the free \Sr\ equation, (\ref{SE(t)}) with $V=0$. 

For the harmonic potential $V= m\omega^2x^2/2$, Senitzky obtained solutions 
\begin{align}\label{SeWF}
	\psi_n(x,t)&\ =\ \left( \frac{\sqrt{m\omega/\pi\hbar}}{2^n\, n!} \right)^{1/2} H_n\left( \sqrt{m\omega/\hbar} \left[x - q(t) \right]  \right)\,e^{-m\omega \left[x - q(t) \right]^2/2\hbar} \cr
	&\  \ \times\ \exp\left\{-\frac{i}{\hbar} \left[ E_n\,t - m \dot{q}(t)\, x + \int \left(\frac 1 2 m\dot{q}^2 - \frac 1 2 m\omega^2 q^2\right) dt   \right]\right\}
\end{align} 
in 1954.  Here $E_n = (n+1/2)\hbar\omega$ and $H_n$ are the Hermite polynomials, for every integer $n\ge 0$.  

$q(t)$ obeys $\ddot q +\omega^2 q = 0$, and therefore describes classical harmonic motion. When $q = 0$, the oscillator energy-eigenstate wave functions $\psi^{\text{HO}}_n(x,t)$ are recovered. Note that the classical harmonic motion described by $q(t)$ has an arbitrary amplitude independent of the width  $\Delta x$ of the $q=0$ wave function.  

 If $n=0$ we get the Gaussian wave function of the well-known coherent state and its dispersionless motion in a harmonic potential. The wave functions involving the excited energy eigenstates with $n>0$ evolve similarly; they are coherent excited states. Since Senitzky's early paper, these solutions have been rediscovered several times, as Philbin recounts~\cite{Ph14}.  

Physically, the Airy beam and coherent excited states are very different, but they have formal similarities. Their wave functions have the same general form 
		\begin{align}\label{Sen}
			\psi(x,t)\ =\ R\big(x-q(t)\big)\, e^{i\varphi(x,t)}\ ;\ \  R,\, \varphi,\,    q(t)\, \in\, {\mathbb R}\ ,  
		\end{align}
that will be explained below. 

The appearance of the Airy function is intriguing.  It recalls the case of a linear potential for which the stationary-state wave functions are Airy functions. Consider $x$ now to be a transformed coordinate, with the original coordinate denoted $x'$.  

For the  Berry-Balazs solution one starts with a linear potential  ${V'} \propto {x'}$  and its energy-eigenstate state Airy-function solutions.  The Airy beam  is obtained by transforming to a uniformly accelerating frame, $x = {x'} + B^3t^2/2$, so that the linear potential transforms to $V=0$, and the force-free solution results.  

Significantly, the transformation just described works on any solution of the Schrödinger equation for a linear potential (see \cite{Va00, Na16, Wa99}, e.g.).  By the equivalence principle, transforming to an accelerating frame of reference has the same consequences as including a uniform gravitational field.  The effect of uniform gravity on quantum mechanics can therefore be calculated using the inverse transformation (see~\cite{Gr80, Na16, CMST22}, e.g.) and has been verified experimentally using neutron interferometry  in the famous ``COW'' experiments \cite{COW75} (see~\cite{GO80} for a popular account). 

We note here that for the Senitzky coherent states an analogous map exists. A harmonic oscillator of frequency $\omega$ can be transformed to a frame oscillating with the frequency $\Omega$ by $x' = x + \beta(t)$, where $\ddot\beta + \Omega^2\beta = 0$.  The resulting classical equation of motion is  
\begin{align}
   \ddot x' \, +\, \omega^2\, x'\, =\, (\omega^2 - \Omega^2)\, \beta\ . 
\end{align} 
If $\Omega=\omega$, however, the fictitious force vanishes. The potential in the new frame has the same form; it is harmonic of frequency $\omega$.  In the quantum case, the transformed  wave functions are Senitzky's (\ref{SeWF}). 

\subsection{General form-preserving tranformations}

A very general framework incorporating the Airy beam and the Senitzky coherent states can be described.  We are interested in transformations $(x, t) \to (x', t')$ that map solutions $\psi(x,t)$ of the \Sr\ equation to other solutions $\psi'(x', t')$, but not necessarily for the same potential. The original and transformed potentials $V$ and $V'$, respectively, need not be equal. That is,  $V'(x'=x, t' = t) = V(x,t)$, is not demanded.

According to~\cite{BS96, Fi99}, the most general transformation of this kind has  
\begin{align}
	{x'}\, =\, \frac{x}{\gamma}\, +\, \beta\, ,\quad {t'}\,& =\, \int^t \,\frac{ds}{\left[ \gamma(s) \right]^{2}}\, , \cr {\psi'}({x'}, {t'})\ =\ \sqrt{\gamma}\,\, \psi(x,t)\,  \exp&\left\{-\frac{i\,m}{2\hbar}\,\left[ \, \frac{\dot \gamma}{\gamma} x^2\, -\, {2\gamma\dot \beta} x\, \right]\, -\, i \, \alpha  \right\} , \cr \frac{{V'}({x'}, {t'})}{\gamma^2}\, =\,  V(x,t)  + \frac{m\ddot \gamma}{2\gamma} {{\, x^2}} -&  m\left(2\dot \gamma\dot \beta + \gamma\ddot \beta\right) {x} +   \frac{m}{2} \gamma^2 \dot \beta^2 + \hbar \, \dot \alpha
\label{form}\end{align}
with $\alpha, \beta$, and $\gamma\not= 0$ arbitrary real functions of $t$.  We will refer to these transformations as ``form-preserving'',  following \cite{Fi99}, or more specifically as \Sr\ form-preserving. 

To verify (\ref{form}), substitute
\begin{align}
    \psi(x,t)\ =\ \sigma(t)\,\, {\psi'}({x'}, {t'})\,\, \exp\left\{\frac i \hbar \phi(x,t) \right\}
\label{sub1}\end{align}
into the time-dependent Schrödinger equation (\ref{SE(t)}), and demand that the same equation  follows for ${\psi'}({x'}, {t'})$ in the primed variables.  Terms proportional to $\partial{\psi'}/\partial{x'}$ must disappear, and the necessary cancellation implies 
\begin{align}
    \frac{\partial\phi}{\partial x}\ =\ -\, m\, \frac{\partial x}{\partial{x'}}\, \frac{\partial{x'}}{\partial t}\ =\ m\left( \frac{\dot\gamma}{\gamma}x\, -\, \gamma\dot\beta \right)\  .
\label{delxphi}\end{align}
Similarly, there are unwanted terms with $(i{\psi'}) e^{i\phi/\hbar}$ multiplied by real quantities. Their removal demands 
\begin{align}
    \frac{\partial^2\phi}{\partial x^2}\ =\ - \frac{2m\dot\sigma}{\sigma}\ .
\label{delx2phi}\end{align}
These equations imply that $\phi$ is quadratic in $x$. From (\ref{delxphi}) we recover 
\begin{align}
\phi(x,t)\ =\ \frac m 2 \left( \frac{\dot\gamma}{\gamma}x^2 - 2\gamma\dot\beta x   \right) \, +\, \hbar \alpha\ ,    
\label{phi1}\end{align}
with $\alpha=\alpha(t)$ arbitrary.  Eqn. (\ref{delx2phi}) relates $\sigma$ to $\gamma$, giving $\sigma = \gamma^{-1/2}$.

Notice that the $\alpha$ part is simply a gauge transformation: if we transform  $\psi\to \psi\, e^{i\alpha}$ so that 
$i\hbar\frac{\partial}{\partial t}\psi\to  i\hbar\frac{\partial}{\partial t}\psi - \hbar\dot\alpha\,\psi$, then $V\to V-\hbar{\dot\alpha}$ compensates and the Schrödinger equation is invariant. 

It is also interesting to re-write the 3rd line of (\ref{form}) as
\begin{align}
	\frac{\partial\phi}{\partial t}\, +\, \frac{1}{2m}\left(\frac{\partial\phi}{\partial x} \right)^2\, +\, \left[ V(x,t) - \frac{{V'}({x'}, {t'})}{\gamma^2}\right]\, =\, 0
\end{align}
using the notation of (\ref{sub1}). We recognize the Hamilton-Jacobi equation for $\phi$ with a ``potential'' $V(x,t) - {{V'}({x'}, {t'})}/{\gamma^2}$. 

The transformation $x\to x'$ in (\ref{form}) is like a Galilean transformation ${x'}=x+vt+b$, ${t'}=t$, extended to include $t$-dependent parameters. It has been called an extended Galilean \cite{GRo72} or quasi-Galilean transformation. When $\beta=0$ and $\gamma$ is independent of $t$, it reduces to the global scaling that leaves the \Sr\ equation invariant. In (\ref{form}), with the time re-parametrization $\gamma =\gamma(t)$, that rescaling is made local in $t$. 

For both the Airy beam and the Senitzky coherent states, $\gamma=1$ and $\beta=-{q}(t)$.  

We find the Berry-Balazs solution when 
\begin{align}
	\alpha \, =\, -\int \Big( m \dot{q}^2/2 + m \ddot{q}\, q \Big) dt\, /\hbar =\, - B^6t^3&/12m^3\hbar\ \ ,\cr \beta\, =\, -{q}\, =\, -B^3t^2/4m^2\  , \ \qquad {\psi'}({x'}, 0)\ =\ \text{Ai}&\left( B{{x'}}/{\hbar^{2/3}}  \right)~.
\label{BBab}\end{align}
Then $V=0$ and $\psi(x,t)$ is the wave function (\ref{BBwf}). 

The Senitzky wave functions (\ref{SeWF}) are produced if 
\begin{align}
	\alpha\, =\,   -\, \int \Big( m \dot{q}^2/2 - m \omega^2 {q}^2/2 \Big) dt\, /\hbar \ ,\qquad \cr \beta\, =\, -{q}\ ,\quad\qquad {\psi'}\left({{x'}}, {{t'}}\right)\ =\ \psi_n^{\text{HO}}\left({{x'}}, {{t'}}\right)~,
\label{Sab}\end{align}
with ${q}(t)$ describing harmonic motion.  In this case, $V = m\omega^2x^2/2$ and ${{V'}} = m\omega^2{{x'}}^2/2$; the original and transformed potentials are identical in form. 

These 2 examples, that led us to (\ref{form}),  only involve Hamiltonians (potentials) that are at most quadratic in position.  The form-preserving transformation (\ref{form}) is not restricted to such cases, however. Finkel et al \cite{Fi99} provide an explicit example with a sextic potential. On the other hand, the difference $V'(x', t')/\gamma^2 - V(x,t)$ is restricted to functions quadratic in $x$, according to (\ref{form}).   

In addition, the 2 cases are special in that $\gamma(t)=1$. For an interesting and useful example with $\gamma(t)\not=1$, consider the map from the harmonic oscillator to the free particle \cite{Ni73, Ta90}. With $V=m\omega^2x^2/2$ and $V'=0$ in (\ref{form}), we require
\begin{align}
    0\ =\ \frac{mx^2}{2} \left(\, \omega^2 + \frac{\ddot\gamma}{\gamma}\, \right)\ +\ mx\left( 2\dot\gamma\dot\beta + \gamma\ddot\beta \right)\ +\ \frac{m\gamma^2\dot\beta^2}2\ +\ \hbar\dot\alpha\ .
\end{align}
Since this must hold for different real $x$, 3 differential equations
\begin{align}
    \ddot\gamma + \omega^2\gamma = 0\, , \qquad \frac 1 \gamma \frac{d}{dt}\left( \gamma^2\dot\beta \right) =0\, ,  \qquad \hbar\dot\alpha + \frac{m\gamma^2\dot\beta^2}2 = 0
\label{threeDEs}\end{align}
must be satisfied. The first tells us that $\gamma$ has harmonic $t$-dependence with frequency $\omega$, so that $\gamma = A\, \cos(\omega t + \varepsilon)$, with amplitude $A$ and phase $\varepsilon$. The second yields $\gamma^2\dot\beta = d\beta/{dt'} = v_0$, a constant. Thus $\beta = v_0 t' + x_0'$ encodes  a constant-velocity translation of the coordinate $x'$. The last equation of (\ref{threeDEs}) gives $\hbar \alpha = -m\left( d\beta/dt' \right)^2/2 = -mv_0^2t'/2$ plus a constant, which we set to 0 since it has no impact on the physics. 

Putting $t'=0$ and $\gamma=1$ at $t=0$ gives $\gamma = \cos(\omega t)$.  Then 
\begin{align}
    t'\ =\ \int^{t} \frac{ds}{\gamma(s)^2}\ =\ \frac 1 \omega\,\tan(\omega t)\ \rightarrow\ \gamma\ =\ \frac 1{\sqrt{1+(\omega t')^2}}\ .
\end{align}
In (\ref{form}), the relation between free and harmonic-oscillator wave functions becomes 
\begin{align}
    \psi\sub{\text{free}}(x', p'; t')\ &=\ \frac 1{\left[ 1+(\omega t')^2 \right]^{1/4}}\, \psi\sub{\text{HO}}\left( \frac{x'- \beta}{\sqrt{ 1 + (\omega t')^2}}, \frac{\tan^{-1}(\omega t')}{\omega}  \right) \cr
    &\ \ \times \exp\left[ \frac{im}{2\hbar}\left( \frac{\omega^2t'(x'-\beta)^2 + 2v_0(x'-\beta)}{1+(\omega t')^2} + v_0^2t' \right)  \right]\ .
\label{psiFreeHO}\end{align}
The free wave function corresponding to any harmonic-oscillator wave function can be found this way.  The free dispersion of any waveform can already be seen from the $t'$-dependent scaling of $x'-\beta$ by $\left[ 1+ (\omega t')^2 \right]^{-1/2}$. 

To see that the time dependence is correct, we now consider the energy-eigenstates of the harmonic oscillator. These wave functions are the only ones that produce time-independent Wigner functions. The time-dependent phase of the wave  function becomes 
\begin{align}
    e^{-iE_nt/\hbar}\, =\, e^{-i(n+\frac 1 2) \tan^{-1}(\omega t')}\, =\, \left[ \frac{\sqrt{1 + (\omega t')^2}}{1+i\omega t'}   \right]^{n+\frac 1 2}\ , 
\end{align}
so that 
\begin{align}
    \psi'(x',t')\ =&\ \frac{\exp\left[ \frac{iv_0}{\omega\ell^2}\left( x'-\frac{v_0t'}{2} \right) \right]}{\sqrt{2^nn!\ell\sqrt{\pi}}   }\, \frac{\left[1 + (\omega t')^2\right]^{\frac n 2}}{\left[1+i\omega t' \right]^{n+\frac 1 2}}    \cr 
    &\times\ \exp\left[ -\frac{(x'-\beta)^2}{ 2\ell^2(1+i\omega t')} \right]\, H_n\left( \frac{x'-\beta}{\ell\sqrt{1+(\omega t')^2}}  \right)\ . 
\label{psindisperse}\end{align}
Here $\ell=\sqrt{\hbar/m\omega}$ is the length scale of the harmonic oscillator (width of the ground state) at $t'=0$. 

Writing $\ell(t') = \ell\sqrt{1+(\omega t')^2}$, the probability density displays the correct structure: 
\begin{align}
    \vert\psi'\vert^2\ =\ \frac{1}{2^nn!\ell(t)\sqrt{\pi}}\, \left\{ H_n\left( \frac{x'-\beta(t')}{\ell(t')} \right)\right\}^2 \,\exp\left\{ -\frac{\left[x'-\beta(t')\right]^2} {\ell(t')^2} \right\}\,\ , 
\end{align}
where $\beta(t') = v_0t'+x_0$. 

We have seen that the Airy beam, the Senitzky coherent excited states, and freely dispersing harmonic waveforms can be understood as arising from form-preserving transformations of the time-dependent \Sr\ equation.  From now on, our focus will be on the form-preserving transformations themselves rather than on particular solutions resulting from them.

\section{\Sr\ form preservation in D space dimensions}

It is natural to ask if the form-preserving transformations work in $D$ space dimensions. The Schrödinger equation is 
\begin{align}
    i\hbar\,\Pl{t}\,\psi\ =\ - \frac{\hbar^2}{2m} \nabla^2 \psi \ + V\, \psi\ , \label{SE(t)3}\end{align}
where now $\psi = \psi( \vec x, t ),\ V = V( \vec x, t )$. 

Let $x_a$ and ${x'}_a$, $a\in \{1,\ldots, D\}$, denote the space coordinates, and $\nabla_a = \Pl{x_a} = \partial_a$, $\nabla'_a = \partial_{x'_a} = \partial'_a$. We use the Einstein summation convention so that ${\nabla}^2 = \partial_{a}\, \partial_{a}$, e.g. 

Consider the transformation
\begin{align}
{x'}_a\ =\ R_{ab}\, &\left(\frac{x_b}{\gamma} + \beta_b \right)\ ,\ \quad {\text{or}}\ \ \ 
{\vec x'}\ =\ R\, \left(\frac{\vec x}{\gamma} + \vec\beta \right)\ ;\cr 
{t'}\,& =\, \int^t \,\frac{ds}{\left[ \gamma(s) \right]^{2}}\ .  
\label{Dzeta}\end{align}
Here $R_{ab}$ are the entries of an orthogonal rotation matrix $R$ (with determinant 1). All of $R$, $\beta$ and $\gamma$ depend on $t$ only. We then have 
\begin{align}
    \Pl{t} =& \gamma^{-2} \Pl{t'} + \Pl{t}\vec x\,'\cdot \Grad' \cr =& \gamma^{-2} \Pl{t'} + \left[ \dot R\left( \frac{\vec x}{\gamma} + \vec\beta \right) + R\left( -\frac{\dot\gamma}{\gamma^2}\vec x + {\dot{\vec\beta}} \right) \right] \cdot \Grad'   
\label{deltp}\end{align}
and 
\begin{align}
    \nabla_a = \gamma^{-1} R_{ba} {\nabla'}_b\quad \ {\text{or}}\quad \ \Grad = \gamma^{-1} R^T\Grad'\ ,    
\label{gradp}\end{align}
so that 
\begin{align}
    \nabla^2 =  \gamma^{-1} R_{ba} \Pl{x'_b}\, \gamma^{-1} R_{ca} \Pl{x'_c}  =   \gamma^{-2}\, {\nabla'}^{\, 2}\ .
\label{laplp}\end{align}

We substitute the $D$-dimensional generalization of ansatz (\ref{sub1}), 
\begin{align}
    \psi(\vec x,t)\ =\ \sigma(t)\,\, {\psi'}({\vec x}^{\,\prime}, {t'})\,\, \exp\left\{\frac i \hbar \phi(\vec x,t) \right\}\ ,
\label{subD}\end{align}
this time into (\ref{SE(t)3}), and demand that the Schrödinger equation for ${\psi'}(\vec x^{\,\prime},t')$ results. Unwanted terms proportional to $\Grad\psi$ are obtained. In order that they cancel, 
\begin{align}
     \Grad\phi = -m\gamma\, R^T\, \Pl{t}{{\vec x}'}\ 
\end{align}
is required. The transformation (\ref{Dzeta}) then yields
\begin{align}
-\frac 1 m\, \Grad\phi\ =\ \left( R^T\dot R - \frac{\dot\gamma}{\gamma} I \right) \vec x + \gamma R^T \dot R\, \vec\beta + \gamma\dot{\vec\beta}\ .
\label{gradphiRdotR}\end{align}
The term $R^T\dot R\, \vec x$ is problematic, however. For a matrix $G$, $\Grad\left(  \vec x^T G\, \vec x\right)\, =\, 2\, G\,\vec x$. But  $R^T\dot R$ is antisymmetric: applying $\frac d{dt}$ to $R^T R =I$, we obtain $R^T\dot R = -( R^T\dot R)^T$. Hence   
\begin{align}
    -\frac m 2\, \vec x^T \,  R^T\dot R\,\, \vec x\ =\ 0\ . 
\label{eRTRdot}\end{align}
We therefore must impose 
\begin{align}
    R^T \dot R = 0 \quad\Rightarrow\quad \dot R = 0\ .
\end{align} 
Since that means $R$ implements an uninteresting global frame rotation, a Galilean symmetry, we put $R=I$. 

Then (\ref{Dzeta}) becomes a simple generalization of the 1-dimensional transformation:
\begin{align}
\vec{x}^{\, '}\ =\ \frac{\vec x}{\gamma} + \vec\beta \ ,\qquad {t'}\,& =\, \int^t \,\frac{ds}{\left[ \gamma(s) \right]^{2}}\ . 
\label{3DRI}
\end{align}
With this simplification (\ref{gradphiRdotR}) can be solved to find
\begin{align}
    \phi\ =\ \frac{m}{2}\left( \frac{\dot\gamma}{\gamma}\, \vec x\cdot\vec x\, -\, 2\gamma \dot{\vec\beta}\cdot {\vec x} \right)\ +\ \hbar\alpha\ ,
\label{phi3}\end{align}
with $\alpha=\alpha(t)$ an arbitrary function of $t$. Compare with (\ref{phi1}). 

Substituting (\ref{subD}) into (\ref{SE(t)3})  gives rise to  further unwanted terms, of the form $i{\psi'}\, e^{i\phi/\hbar}$ times a real quantity. Such terms also appeared in the 1D case of Section 2. Here their elimination requires $\nabla^2\phi = -2m \dot\sigma/\sigma$. Using (\ref{phi3}) then gives $2\dot\sigma/\sigma = -D\dot\gamma/\gamma$, yielding $\sigma = \gamma^{-D/2}$. The form-preserving transformations go through in D-dimensions with this change.  

With the unwanted terms eliminated, substituting (\ref{3DRI}) into (\ref{SE(t)3}) does result in the Schrödinger equation for ${\psi'}( {\vec x}\, ',  t')$, so that the relation 
\begin{align}
    \frac{ V'( {\vec x}\, ',  t' )}{\gamma^2}\ =\ V(\vec x, t)\ +\ \frac{\pl\phi(\vec x, t)}{\pl t}\ +\ \frac{\left[ \nabla\phi(\vec x, t) \right]^2}{2m}
\label{VolV}\end{align}
can be read off.  If needed, the particular expression (\ref{phi3}) can be substituted into this general formula. 

As in the 1-dimensional case, time-dependent translations and time-dependent scaling can be done \cite{GRo72}, but time-dependent rotations seem to be excluded. 

However, if one includes a vector potential $\vec A$ as well as the scalar potential $V$, form preservation involving rotations can be realized \cite{Nikitin2020}. It becomes possible because with $\vec A$ present the \Sr\ equation must include a term that is 1st-order in spatial derivatives. Form preservation can then result when both the scalar and vector potentials transform non-trivially. 

The simplest \Sr\ equation with both  scalar and vector potentials is 
\begin{equation}
    \left[\frac{1}{2m}\left(-i\hbar\Grad - \vec A\right)^2 + V\right]\psi = i\hbar\, \Pl{t}{\psi}\ . 
\label{U1SE}\end{equation}
Because it is invariant under the $\mathrm{U(1)}$ gauge transformation 
\begin{align}
    \vec A \to \vec A+\Grad{\Lambda}\ ,\quad V \to V-\pdv{\Lambda}{t}\ ,\quad   \psi\to \psi\, e^{i\Lambda/\hbar}\ , 
\label{U1SE1}\end{align}
it can be called the $\mathrm{U(1)}$ \Sr\ equation. 

Transformations of the equation (\ref{U1SE}) were investigated by Takagi in \cite{TaI91}, where it was argued that (\ref{Dzeta}) describes the most general possible form-preserving transformation.\footnote{One of us has also presented an argument for the generality of (\ref{Dzeta}) in \cite{Daub25}, where other results related to this work can be found.} In \cite{Nikitin2020} Nikitin independently classifies the symmetries of (\ref{U1SE}) for different specified vector and scalar potentials, and also investigates form-preserving transformations, calling them equivalence transformations. We will contribute here by describing the transformations in this case explicitly and generally, in the manner of (\ref{form}) \cite{Fi99}: see equations (\ref{Dformpsi}, \ref{DformA}, \ref{VdimD}) below. 

Expanding the square of \eqref{U1SE} yields 
\begin{equation}
\label{U1SE2}
    -\frac{\hbar^2}{2m}\nabla^2 \psi+\frac{i\hbar}{2m}\vec A\cdot \Grad{\psi} +\left(\frac{i\hbar\Grad\cdot{\vec A}+\vec A^{\, 2}}{2m}+V\right)\psi = i\hbar \,\Pl{t}{\psi}\ .
\end{equation}
We now use the ansatz (\ref{Dzeta}), assume that  $\psi$ satisfies
\eqref{U1SE2} and we demand that $\psi'$ obey the same equation in the primed coordinates with vector potential $\vec A'$ and scalar potential $V'$. 

Inserting (\ref{subD}) into (\ref{U1SE2}) and using (\ref{deltp}, \ref{gradp}, \ref{laplp}),  we find
\begin{multline}
    -\frac{\hbar^2}{2m\gamma^2}\nabla'\, ^2 \psi' + \frac{i\hbar}{m}\left[ -\frac{\Grad\phi}{\gamma} + \frac{\vec A}\gamma - mR^T \frac{\del\vec x\, '}{\del t}\right]\cdot \left(R^T\Grad'\psi'\right) + \\ 
    \left[ -\frac{i\hbar\nabla^2\phi}{2m} + \frac{(\nabla\phi)^2}{2m} - \frac{\vec A\cdot \Grad\phi}{m} + {\frac{i\hbar\Grad\cdot\vec A + \vec A^2}{2m} + V - \frac{i\hbar\dot\sigma}{\sigma} + \frac{\del\phi}{\del t}}\right] \psi'   \\  = \frac{i\hbar}{\gamma^2} \frac{\del\psi'}{\del t'}\ . \qquad\qquad\qquad{} 
    \label{sub3}
\end{multline}
Consider the $\Grad{\,}'\psi'$ terms. The corresponding term was required to vanish when no vector potential was present. Here, however, we see that the coefficient of $R^T\,\Grad\, '\psi\, '$ must be the rotated and scaled vector potential $\gamma^{-2} R^T \vec A{\,}'$. This implies 
\begin{equation}
\label{eqn:3dphase}
\begin{split}
     \Grad \phi &= -m\gamma\, R^T\,\Pl{t}{\vec x\, '} + \vec A - \frac{1}{\gamma}R^T\vec A\, '\\
    &=m\left(\frac{\dot\gamma}{\gamma}\vec x-\gamma\dot{\vec \beta}\right) + \vec A - \frac{1}{\gamma} R^T\vec A' - m R^T\dot{R}(\vec x+\gamma\vec\beta).
\end{split}
\end{equation}

Consider the $R^T \dot{R} \vec{x}$ term. Recall that the matrix $R^T \dot{R}$ is skew-symmetric for all orthogonal matrices $R$, and that the equation $\Grad \Phi= M\vec x$ has no solution for a skew-symmetric matrix $M$, since $x_a M_{ab} x_b = 0$. The vector potentials themselves are arbitrary and cannot in general be written as the gradient of a scalar field.

Therefore, in order for $\phi$ to exist, we take the transformed vector potential to be
\begin{equation}
    \frac{ R^T\vec A\, '}{\gamma} = \vec A -\Grad \alpha - m R^T\dot{ R}(\vec x + \gamma\vec\beta)\ .
\label{A'A}\end{equation}
Here $\alpha(\vec x, t)$ is a real, arbitrary scalar field, which represents the freedom to choose the gauge of both $\vec A\, '$ and $\vec A$ when we perform the coordinate transformation. It is easier to see that $\alpha$ can also be a gauge transform of $\vec A\, '$ if we remember that $\gamma R\,\Grad\alpha = \Grad\,'\alpha$. We also use the arbitrariness of $\alpha$ to keep the combination $\vec x + \gamma\vec\beta$ intact. The phase can now be written as 
\begin{equation}
     \phi(\vec x, t) = \frac{m}{2}\left(\frac{\dot\gamma}{\gamma}\vec x^2 - 2\gamma\dot{\vec\beta}\cdot\vec x\right) + \alpha(\vec x, t)\ .
\label{phialpha}\end{equation}

This leaves us with the Schr\"odinger equation
\begin{multline}
\label{eqn:3dcase-step4}
     \gamma^2\bigg[-\frac{i\hbar}{2m}\nabla^2 \phi +\frac{(\nabla \phi)^2}{2m}-\frac{1}{m}\vec A\cdot \Grad\phi  + \frac{i\hbar\Grad\cdot{\vec A}+\vec A^2}{2m}+V  \\ -\frac{i\hbar\dot \sigma}{2\sigma} +  \pdv{\phi}{t}\bigg]\psi' 
     -\frac{\hbar^2}{2m}\nabla'^2 \psi' + \frac{i\hbar}{m}\vec A'\cdot \nabla'\psi'=i\hbar \pdv{\psi'}{t'}.
\end{multline}
Upon comparison with the transformed equation, we identify the coefficient of $\psi'$ with
\begin{equation}
    \frac{i\hbar \Grad\,'\cdot\vec A\,' + \vec A\,'^{\, 2}}{2m} + V'.
\end{equation}
We therefore have
\begin{align}
    -\frac{i\hbar}{2m}\nabla^2 \phi\, +\, &\frac{(\nabla \phi)^2}{2m}-\frac{1}{m}\vec A\cdot \Grad\phi +    \frac{i\hbar\Grad\cdot{\vec A}+\vec A^{\, 2}}{2m} + V - \frac{i\hbar\dot \sigma}{2\sigma} + \pdv{\phi}{t} \cr  &= \frac{i\hbar \Grad'\cdot\vec A\,' + \vec A\,'{\, ^2}}{2m\gamma^2} + \frac{V'}{\gamma^2}.
\end{align}
By (\ref{phialpha}) and (\ref{A'A}) the Laplacian of the phase is 
\begin{align}
 \nabla^2\phi\, =\, \frac{m\dot\gamma}{\gamma}\,\Tr I\, +\, \Grad\cdot\vec A\, -\, \frac 1{\gamma^2}\,\Grad\,'\cdot\vec A\,'\, +\, \nabla^2\alpha\ .
\end{align}
The trace of the identity is just the number of dimensions, $D$. From (\ref{A'A}) we find 
\begin{align}
\frac 1{\gamma^2}\, \Grad\,'\cdot\vec A\,'\ =\, \Grad\cdot\vec A\, -\, \nabla^2\alpha\ ,     
\end{align}
where we used that the trace of a skew-symmetric matrix like $R^T\dot R$ vanishes. We are left with
\begin{equation}
    \frac{1}{\gamma^2}\left(V' +\frac{\vec A\,'^{\, 2}}{2m}\right)= V + \frac{\partial\phi}{\partial t}  + \frac 1{2m}\,\left(\vec A - \Grad\phi\right)^2 - {i\hbar }\left( \frac{D\dot\gamma}{2\gamma}+\frac{\dot \sigma}{\sigma}\right).
\end{equation}
Only the term involving $\gamma, \sigma$ is imaginary, and so with 
\begin{equation}
    \sigma = \gamma^{-\frac D 2}\     
\end{equation}
vanishes. 

Overall, this means that original and transformed wave functions are related by 
\begin{equation}
    \psi'(\vec x\,', t\,') = \gamma^{D/2}\,\,\psi(\vec x, t) \,\, \exp\left[-\frac{im}{2\hbar}\left(\frac{\dot\gamma}{\gamma}\vec x^{\, 2} - 2\gamma\dot{\vec\beta}\cdot \vec x\right) - \frac{i\alpha(\vec x, t)}{\hbar}\right],
\label{Dformpsi}\end{equation}
where $\vec x\,'$ and $t\,'$ are determined by the coordinate transformation (\ref{Dzeta}). The transformed wave function satisfies the D-dimensional $\mathrm{U(1)}$ \Sr\ equation in the primed coordinates with a vector potential
\begin{equation}
    \vec A' = \gamma R(\vec A - \Grad\alpha) - \gamma m\dot{ R}(\vec x + \gamma\vec\beta)
\label{DformA}\end{equation}
and a scalar potential
\begin{equation}
        \frac{V\,'}{\gamma^2} = V + \pdv{\phi}{t} + \frac{(\vec A -\Grad\phi)^2}{2m} - \frac{(\vec A\,'/\gamma)^2}{2m}\ .
\label{VdimD}\end{equation}
It is interesting to rewrite this as 
\begin{align}
    \frac 1{\gamma^2}\, \left(  V\,'\, +\, \frac{\vec A\,'^{\,\, 2}}{2m} \right)\ =\ V\, -\, \pdv{\Lambda}{t}\, +\, \frac{(\vec A\, +\, \Grad\Lambda)^2}{2m}
\end{align}
with $\Lambda = -\phi.$  Comparing with (\ref{U1SE1}), we see that the combination $ V + \vec A^{\,\, 2}/2m$ transforms simply, by a gauge transformation generated by $-\phi$ followed by a scaling by $\gamma^2$.

\subsection{\Sr\ form preservation in 3 dimensions}
In this subsection we  focus on the special case of $D=3$ spatial dimensions, and matrices $R$ in the $\mathrm{SO(3)}$ subgroup of $\mathrm{O(3)}$. To start, the $\dot R$ terms in the transformations (\ref{DformA}, \ref{VdimD}) will be rewritten in familiar $D=3$ vector-product notation.  

When $D=3$, the Levi-Civita tensor defines the vector (or cross) product:
\begin{align}
   \vec a\ =\ \vec b \, \times\, \vec c\ \ \ \to\ \ \ a_i\ =\ \epsilon_{ijk}\, b_j\, c_k\ .  
\end{align}
Now $\epsilon_{ijk}$ transforms as an $\mathrm{SO(3)}$-invariant tensor:  
\begin{align}
    \epsilon^{\, '}_{ijk}\ =\ R_{i\ell} R_{jm} R_{kn}\, \epsilon_{\ell m n}\ =\ \text{det}(R)\, \epsilon_{ijk}\ =\ \epsilon_{ijk}\ , 
\end{align}
if $R\in \mathrm{SO(3)}$, so that $\text{det}(R)=+1$. That implies 
\begin{align}
    (R\vec a) \, \times\, (R\vec b)\ =\ R(\vec a \, \times\, \vec b)\ . 
\label{Rcross}\end{align}
The Levi-Civita symbol also allows the entries of any skew-symmetric $3\times 3$ matrix $A$ (such as $R^T\dot R$) to be encoded in a (pseudo-)vector $\vec s$:  $A_{ij} = \epsilon_{ikj}\, s_k$. Therefore we put $(R^T\dot R)_{ij}= \epsilon_{ikj}\, \omega_k$, so that  
\begin{align}
     \dot R\, \vec u \ =\ R \left( \vec\omega \times \vec u \right)\ ,
\label{Rdotu}\end{align}
for any vector $u$. 

The pseudo-vector $\vec\omega$ is the instantaneous angular velocity of the time-dependent frame rotation $R(t)$. To see why, consider a fixed point $\vec x$ in the nonrotated coordinates. The corresponding point in the rotated coordinates is $\vec x\,'= R\,\vec x$. The velocity of this point in the rotating frame is the time derivative $d{\vec x}\,'/dt = \dot{ R}\, \vec x$, which must equal  $\vec \omega\,' \cross \vec x\,'$, where $\vec \omega\,' = R\,\vec\omega$ is the instantaneous angular velocity in the rotated coordinates. So by (\ref{Rcross}), we must have $ \dot R\, \vec x \ =\ R \left( \vec\omega \times \vec x \right)$. Comparing this with  (\ref{Rdotu}) yields the desired identification of $\vec\omega$. 

Thus we can write the transformed vector potential (\ref{DformA}) in 3-dimensions as 
\begin{equation}
\begin{split}
    \vec A\,' &= \gamma R (\vec A-\Grad \alpha) -m\gamma R[\vec\omega \cross(\vec x+\gamma\vec\beta)]\\
    & = \gamma R \vec A-\Grad\,' \alpha -m\gamma^2( R\vec\omega)\cross\vec x\,'.
\end{split}
\label{ApA}\end{equation}
Now that we are in 3 dimensions we can also define the magnetic field as the curl of the vector potential. The transformed magnetic field should then be the curl in the primed coordinates of the new vector potential, or $\vec B' =\Grad\,'\cross\vec A'$. Taking the curl of (\ref{ApA}) gives the transformed magnetic field
\begin{equation}
    \vec B' = \gamma^2  R(\vec B - 2m\vec\omega),
\end{equation}
where $\vec B = \Grad \times {\vec A}$ is the magnetic field before the transformation. Finally,  the transformed scalar potential (\ref{VdimD}) is
\begin{multline}
\label{eqn:3dpotential}
    \frac{V'}{\gamma^2} = V + \pdv{\alpha}{t} +\frac{m\ddot\gamma}{2\gamma}\vec x^2-m\vec x\cdot(2\dot\gamma\dot{\vec\beta}+ \gamma\ddot{\vec\beta}) + \frac{m}{2}(\gamma\dot{\vec\beta})^2\\ - \frac{m}{2}[\vec \omega\cross(\vec x+\gamma\vec\beta)]^2 - (\vec A - \Grad\alpha)\cdot\left[\frac{\dot\gamma}{\gamma}\vec x - \gamma\dot{\vec \beta} - \vec\omega \cross (\vec x+\gamma\vec\beta)\right].
\end{multline}
Compare this result with that obtained for $D=1$, equation (\ref{form}).  The top line is just the $D=3$ version of the $D=1$ result. The 1st of the 2 additional terms on the 2nd line is the kinetic energy of the rotating frame which creates the outwards centrifugal force.  The second term is the interaction potential of the moving frame with the magnetic field.

\section{Form-preserving transformations in phase space}

Form-preserving transformations reveal new properties when considered in phase space. We will first consider the effect of \Sr\ form-preserving transformations on Wigner functions. A strikingly simple result is obtained.  We show how it explains and generalizes the rigid translation of phase-space curves related to the Airy beam and Senitzky coherent excited state wave functions. Then we consider the Moyal equation of motion obeyed by the Wigner function in phase space to derive its form-preserving transformations directly. In this section we restrict attention to $D=1$-dimensional case. 
 
\subsection{Wigner function transformation from \Sr\ form preservation}
Phase space can host quantum mechanics, but a true quantum probability density is impossible there. Although a classical state of a fixed position and momentum can exist, such a quantum state cannot because of the uncertainty principle. The Wigner function $W(x, p; t)$ can be used instead (see \cite{CFZ14, ZFC05, Ha04}, e.g., for pedagogical treatments). It is analogous to but not a true probability density; it is a quasi-probability distribution. That is clear because it can be (and almost always is) negative in regions of phase space. 

The Wigner function can be calculated from the wave function: 
\begin{align}
W(x,p;t)\ =\ \frac 1 {\pi\hbar}\, \int_{-\infty}^\infty dy\, e^{2ipy/\hbar}\, \psi(x-y,t)\,\psi^*(x+y,t)\ .
\label{WigPsi}\end{align}
The transformation (\ref{form}) can be extended to phase space as 
\begin{align}
    x'\ =&\ x/\gamma +\beta\ \ ,\cr
    {p'}\ =&\ m\frac{d{x'}}{d{t'}}\ =\ \gamma p\, +\, m\gamma^2\dot\beta\, -\, m{\dot\gamma} x\, \ .
\label{CT}\end{align} 
The Jacobian of this transformation is 1, so that (\ref{CT}) is a linear canonical transformation.

If the Wigner function ${W'}({x'}, {p'};{t'})$ is calculated from ${\psi'}({x'},{t'})$ as in (\ref{WigPsi}), it is straightforward to show that the simple result 
\begin{align}
    {W'}({x'}, {p'}; {t'})\ =\ W(x,p; t)\  
\label{WolW}\end{align}
holds. 

This result describes one more way the Wigner function, a quasi-probability density,  behaves like a true, measurable probability density. To see this, consider a probability density $\Phi$ on a space $\Xi$, and the probability $\Phi(\xi)\, d\xi$ of measuring a value $\xi$ in a neighbourhood of the point $\xi\in \Xi$. Here $d\xi$ denotes a volume element containing $\xi\in\Xi$. Since a change of coordinates won't change that probability, we have 
\begin{align}
   \Phi(\xi)\, d\xi\ =\ \Phi'(\xi')\, d\xi'\ \ . 
\end{align}
If the determinant of the Jacobian for the transformation $\xi\to \xi'$ is 1, however, $d\xi = d\xi'$, so that $\Phi(\xi) = \Phi'(\xi')$ follows. Since the transformation (\ref{CT}) is canonical, it fits this general description, and (\ref{WolW}) holds. 

The result (\ref{WolW}) explains a remarkable property of the wave functions that were discussed above. Berry and Balazs \cite{BB79} show that a parabola in phase space is accelerated rigidly in the flow generated by the free Hamiltonian. They considered the phenomenon a signal of the existence of their Airy beam solution. Confirmation was produced in \cite{MA19}, where the corresponding Wigner function was shown to have parabolic level curves that accelerated rigidly. Here we demonstrate that the rigid acceleration follows from (\ref{WolW}). 

Suppose that the wave function ${\psi'}$ describes a stationary state, so that ${W'}$ is independent of time: $W'(x', p'; t') = W'(x', p'; 0)$.  We then have
\begin{align} 
    W(x,p; t)\ =\ W'(\, x/\gamma +\beta,\, \gamma p\, +\, m\gamma^2\dot\beta\, -\, m{\dot\gamma} x; 0\,)\ .
\label{genrigid}\end{align}
If we specialize to $\gamma=1$ the result (\ref{genrigid}) becomes 
\begin{align} 
    W(x,p; t)\ =\ W'\left(\, x + \beta,\,  p\, +\, m\dot\beta; 0 \,\right)\ .
\label{gonerigid}\end{align}
As time passes, the Wigner function is translated rigidly in phase space. 

For the Berry-Balazs solution, $\gamma = 1$, $\beta\, =\,  -B^3t^2/4m^2$ and $\psi'$ is the energy eigenstate wave function of Airy-function form (\ref{BBab}). The Wigner-function level curves are accelerating parabolas in phase space, depicted in Figure 1.

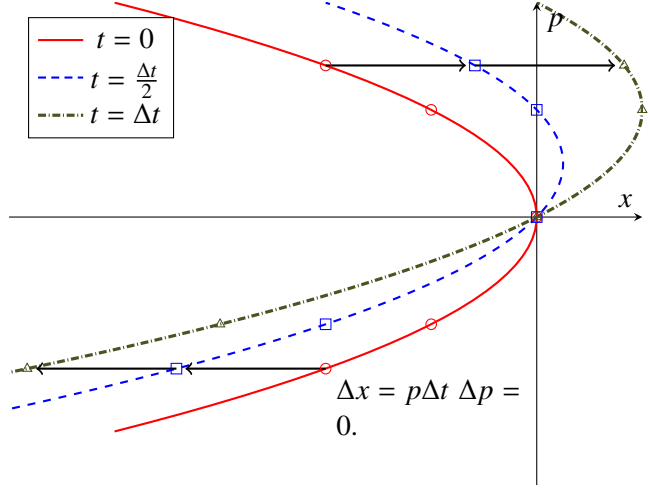
\begin{figure}
    \centering
    \begin{tikzpicture}
        \begin{axis}[xlabel=$x$, ylabel=$p$, axis lines = middle, xtick = {0}, 
            ytick= {0},ymin = -2.5,xmin=-2.5,xmax=0.5,legend style={legend pos = north west},
            width =10cm, height =8cm]
            \addlegendimage{red,  thick};  %mark=o,
            \addlegendentry{$t=0$};  %\addlegendentry{$t=\SI{0}{s}$};
            \addlegendimage{blue,  densely dashed, thick};   %mark=square,
            \addlegendentry{$t=\frac{\Delta t}{2}$};   %\addlegendentry{$t=\SI{0.5}{s}$};
            \addlegendimage{armygreen,  densely dash dot, very thick};  %mark=triangle,
            \addlegendentry{$t=\Delta t$}

            % Red Parabola
            \addplot[red, thick, domain=-2:2, samples = 50] ({-(x)^2/2},{x}); 
            \addplot[only marks, mark=o, red] coordinates {(-1, -1.414) (-0.5, -1) (0,0) (-0.5, 1) (-1, 1.414)};

            % Blue Parabola
            \addplot[blue, dashed, thick, domain=-2:2, samples=100] ({-(x-0.5)^2/2+0.125},{x}); 
            \addplot[only marks, mark=square, blue] coordinates {(-1-1.414*0.5, -1.414) (-0.5-1*0.5, -1) (0,0) (-0.5+1*0.5, 1) (-1+1.414*0.5, 1.414)};

            % Orange Parabola
            \addplot[armygreen, densely dash dot, very thick, domain=-2:2, samples = 100] ({-(x-1)^2/2+0.5},{x});
            \addplot[armygreen, only marks, mark=triangle] coordinates {(-1-1.414*1, -1.414) (-0.5-1*1, -1) (0,0) (-0.5+1*1, 1) (-1+1.414*1, 1.414)};

            % Movement Arrows
            \draw[->,thick] (-1,-1.414) node[anchor = north west, text width = 2.5cm] {$\Delta{x}=p\Delta{t}$ $\Delta{p}=0.$} -- (-1 - 0.47 * 1.414, -1.414) ;
            \draw[->, thick] (-1-1.414*0.5, -1.414) -- (-1-0.97 * 1.414,-1.414);
            \draw[->, thick] (-1, 1.414) -- (-1 +0.47 * 1.414, 1.4141);%-- (0.414, 1.414);
            \draw[->,thick] (-1 +0.5*1.414, 1.414) -- (-1 + 0.97 * 1.414, 1.414);
            
        \end{axis}
    \end{tikzpicture}
    \caption{A phase-space parabola evolved rigidly in time by the free Hamiltonian. It relates to the free-space accelerating Airy beam solution of Berry and Balazs \cite{BB79}, being a level curve (contour line) of the corresponding Wigner function \cite{MA19}. } 
    \label{fig:FSrigid}
\end{figure}

\begin{figure}
    \centering
    \begin{tikzpicture}
        \begin{axis}[xlabel=$\tilde x$, ylabel=$\tilde p$, axis lines = middle, xtick = {0}, 
            ytick= {0},ymin = -2., ymax=2., xmin=-2.,xmax=2.,legend style={legend pos = north west},
            width =8.5cm, height =8.5cm]
            \addlegendimage{red, thick};
            \addlegendentry{$t=0$};  %\addlegendentry{$t=\SI{0}{s}$};
            \addlegendimage{blue, dashed, thick};
            \addlegendentry{$t=\Delta t/2$};   %\addlegendentry{$t=\SI{0.5}{s}$};
            \addlegendimage{armygreen, densely dash dot, very thick};
            \addlegendentry{$t=\Delta t$}

             \draw[gray, fill=none, loosely dashed](0,0) circle (1.25); %node [blue] 

            % Red Circle
            %\addplot[red, thick, domain=-2:2, samples = 50] ({-(x)^2/2},{x}); 
            \draw[red, thick, fill=none](0.8034845, 0.957556) circle (0.4); %node [blue] 
            \addplot[only marks, thick, mark=o, red] coordinates {(0.546369, 0.651138) (0.4970667, 1.2146706) (1.0605996, 1.2639733) (1.1099023, 0.7004405)};

            % Blue Circle
            %\addplot[blue, thick, domain=-2:2, samples=100] ({x}, {x}); 
            \draw[blue, thick, dashed, fill=none](1.25, 0) circle (0.4); %node [blue] 
            \addplot[only marks, thick, mark=square, blue] coordinates {(1.25, 0.4) (1.25, -0.4) (1.25-0.4, 0) (1.25+0.4, 0)};

            % Orange Circle
            %\addplot[orange, thick, domain=-2:2, samples = 100] ({-(x-1)^2/2+0.5},{x});
             \draw[armygreen, very thick, densely dash dot, fill=none](0.8034845, -0.957556) circle (0.4); %node [blue] 
            \addplot[armygreen, thick, only marks, mark=triangle] coordinates {(0.546369, -0.651138) (0.4970667, -1.2146706) (1.0605996, -1.2639733) (1.1099023, -0.7004405)};

        \end{axis}
    \end{tikzpicture}
    \caption{A phase-space circle evolved rigidly in time by the simple harmonic  Hamiltonian. The Senitzky solutions describing coherent excited states \cite{Se54} have Wigner functions with level curves that behave this way. } 
    \label{fig:SenRigid}
\end{figure}
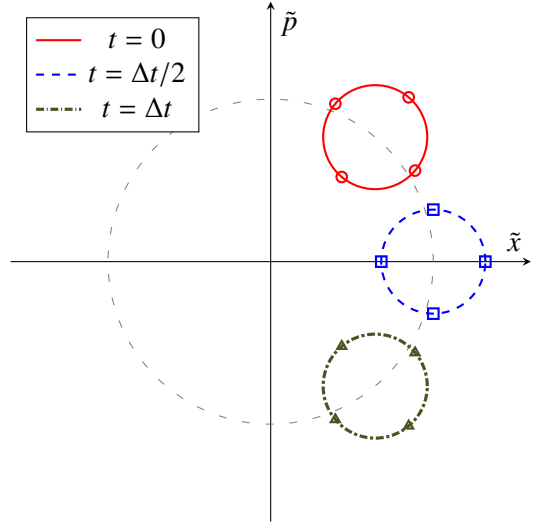

Further examples with $\gamma = 1$ are furnished by the Senitzky coherent excited states.  The Wigner function transformation in this case is described by  $\beta(t)$ harmonic, i.e. $\ddot\beta + \omega^2\beta = 0$, and  $\psi'$ an energy eigenstate wave function (\ref{Sab}). Rewritten in terms of dimensionless position  $\tilde x =$ $x/\ell$ $=x\sqrt{m\omega/\hbar}$ and dimensionless momentum $\tilde p =$ $p\,\ell/\hbar =$ $p/\sqrt{\hbar m\omega}$, the Wigner-function level curves are circles. Their centre translates around a second circle centred on the origin. Its radius is arbitrary, equal to the amplitude of $\beta(t)$.  This arrangement is illustrated in Figure \ref{fig:SenRigid}.

This rigid evolution (\ref{gonerigid}) in the case of of the Senitzky Wigner functions is also a special example of the harmonic ``turntable'' flow of Wigner functions described in Figure 4 of \cite{ZFC05} and Figure 6 of \cite{CFZ14}. To show this, introduce 
\begin{align}
    \zeta\ =\ \left(\begin{matrix} x \cr p\end{matrix}\right)\ ,
\label{defzeta}\end{align}
its dimensionless version $\tilde\zeta$, and $\tilde t =\omega t$.  Writing $W(x, p; t) = \tilde W(\tilde\zeta;\tilde t)$, the turntable evolution is 
\begin{align}
  \tilde W(\tilde\zeta;\tilde t)\ =\  \tilde W( R(\tilde t)\,\tilde\zeta; 0), 
\label{turntable}\end{align}
where $R(\theta)$ is the rotation matrix 
  \begin{align}
      R(\theta)\ =\ \left(
      \begin{matrix}
         \cos\theta  &   -\sin\theta  \cr
          \sin\theta  &   \cos\theta
      \end{matrix} \right)\ .
  \end{align}
But for the Senitzky Wigner functions,  the rigid translation (\ref{gonerigid}) can be written as  
\begin{align}
   \tilde W(\tilde\zeta;\tilde t)\ =\  \tilde W'( \tilde\zeta\, +\, R^{-1}(\tilde t)\,\tilde\zeta\sub{0}; 0)\ ,  
\label{rigidzeta}\end{align}
where $\tilde\zeta\sub{0}$ indicates a fixed point in phase space. This implies that
\begin{align}
    \tilde W(\tilde\zeta; 0)\ =\  \tilde W'( \tilde\zeta\, +\, \tilde\zeta\sub{0}; 0)\ . 
\label{zetazero}\end{align}
As is clearly displayed in the figures just mentioned, the rigid translation can only be equivalent to turntable flow (rotation about the origin) if the Wigner function has circular symmetry. Harmonic oscillator stationary states are circularly symmetric; they obey 
\begin{align}
    \tilde W'(\tilde\zeta; 0)\ =\ \tilde W'( R_\theta\,\tilde\zeta; 0)\ ,  
\end{align}
for any $\theta$. Applying this with $\theta=\tilde t$  to (\ref{rigidzeta}) yields 
\begin{align}
   \tilde W(\tilde\zeta;\tilde t)\ =\  \tilde W'( R(\tilde t)\,\tilde\zeta\, +\, \tilde\zeta\sub{0}; 0)\ .  
\end{align}
Because of (\ref{zetazero}), the turntable flow (\ref{turntable}) follows. 

In the special case of $\gamma=1$, the Wigner form-preserving transformation (\ref{WolW}) explains the Wigner rigid translation observed for the Airy beam and the coherent excited states. It also generalizes the behaviour to $\gamma\not=1$. 

A notable example of the generalization (\ref{genrigid}) with $\gamma \not= 1$ is the free-harmonic  equivalence \cite{Ni73, Ta90}. Included is the map from a harmonic oscillator stationary state to its dispersing free state. 
Going back to the notation of (\ref{psiFreeHO}), we write 
\begin{align}
  W'_{\text{free}}(x', p'; t')\ =\ W_{\text{HO}}(x, p; 0)\  .  
\label{freeHOi}\end{align}
The argument $t$ is put to zero, assuming the state is stationary. For the $n$-th stationary state of  energy $E=(n+1/2)\hbar\omega$ the Wigner function is (see \cite{CFZ14, ZFC05}, e.g.) 
\begin{align}
 W_{\text{HO},\, n}(x,p)\ =\ \frac{(-1)^n}{\pi\hbar}\, L_n\left( \frac{4H}{\hbar\omega} \right)\, \exp\left\{ - \frac{2H}{\hbar\omega} \right\}\ ,  
\end{align}  
where $H$ is the simple-harmonic Hamiltonian and  $L_n$ is the $n$-th Laguerre polynomial. Note that 
\begin{align}
    \frac{2H}{\hbar\omega}\ =\ \tilde x^2\ +\ \tilde p^2\ .
\label{Hcircle}\end{align}
The level curves of the stationary harmonic oscillator Wigner functions are circles in these variables.

From (\ref{CT}) we have 
\begin{align}
    x\, =\, \gamma\,\left( x' - \beta \right)\ ,\ \qquad  
    {p}\, =\, \frac 1 \gamma\, p'\, -\, m\,\gamma\,\dot\beta\, +\, m\,x'\, {\dot\gamma}\, \ .
\label{CTinv}\end{align} 
Here $\gamma = \left[ 1+ (\omega t')^2\right]^{-1/2}$ and we put $\beta = x_0 + v_0t = 0$ to focus on the dispersion.  Note that 
\begin{align}
    \dot\gamma\ =\ \frac{d\gamma}{dt}\ =\ \frac{dt'}{dt}\, \frac{d\gamma}{dt'}\ =\ -\frac{\omega^2 t'}{\sqrt{1+(\omega t')^2}}\ . 
\end{align}
Then (\ref{freeHOi}) becomes
\begin{align} 
    W_{\text{free},\, n}(x',p'; t')\ =&\ \\ W_{\text{HO},\, n}\Bigg(\, \frac{x'}{\sqrt{1 + (\omega t')^2}} ,&\,\, p'\, \sqrt{1 + (\omega t')^2}\,  -\, m{\omega} x'\,   \frac{\omega t'} {\sqrt{1+(\omega t')^2}}\,\Bigg)\ . 
\label{Wndisperse}\end{align}
Finally, we drop the primes, and 
\begin{align} 
    W_{\text{free},\, n}(x, p; t)\ =\ \frac{(-1)^n}{\pi\hbar}\, L_n\left( \frac{4H(x, p; t)}{\hbar\omega} \right)\, \exp\left\{ - \frac{2H(x, p; t)}{\hbar\omega} \right\}\ ,  
\label{Wndispnoprimes}\end{align} 
where (\ref{Hcircle}) is replaced by 
\begin{align}
    \frac{2H(x, p; t)}{\hbar\omega}\ =\ \tilde x^{\, 2}\ -\ (\, 2\,\tilde t\,)\,\tilde x\,\tilde p\ +\ (1\, +\, \tilde t^{\, 2})\,\tilde p^{\, 2} \ .
\label{Hcirclet}\end{align} 

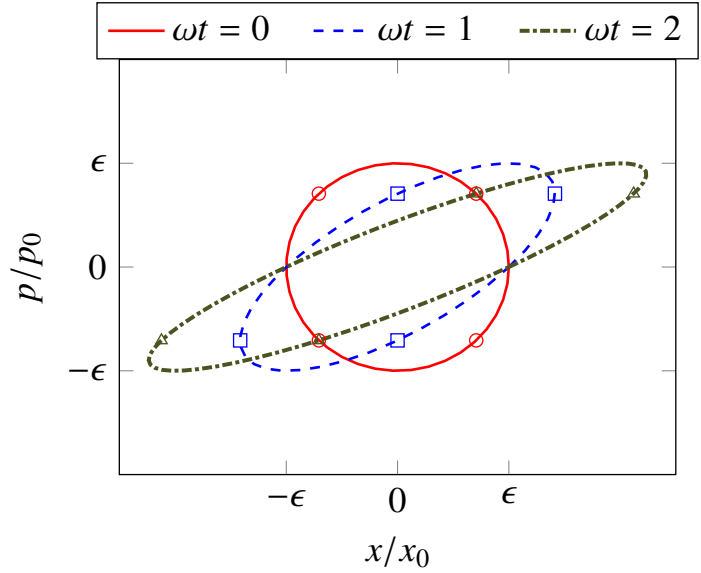
\begin{figure}
    \centering
    \begin{tikzpicture}[scale=1.25]
        \begin{axis}[
            xmin =-2.5, 
            xmax = 2.5,
            ymin = -2,
            ymax = 2,
            width = 7.5cm,
            height = 6cm,
            xtick={-1, 0,1},
            xticklabels={$-\epsilon$, 0,$\epsilon$},
            ytick={-1, 0,1},
            yticklabels={$-\epsilon$, 0,$\epsilon$},
            xlabel={$x/x_0$},
            ylabel={$p/p_0$},
            legend columns = 3,
            legend style = {at={(0.5, 1)},anchor = south}
        ]
        \addplot [red, thick, domain=0:360, samples=40] 
        ({cos(\x)}, {sin(\x)} );
        \addplot [blue, thick, dashed, domain=0:360, samples=40] 
        ({cos(\x)+sin(\x)}, {sin(\x)} );
        \addplot [armygreen, very thick, densely dash dot, domain=0:360, samples=60] 
        ({cos(\x)+2*sin(\x)}, {sin(\x)} );
        \legend{{$\omega t = 0$\quad${}$},{$\omega t = 1$\quad${}$},{$\omega t = 2$}};
        
        \addplot[red, only marks, mark = o] coordinates{(0.707, 0.707) (-0.707,0.707) (-0.707,-0.707) (0.707,-0.707)};
        \addplot[blue, only marks, mark = square] coordinates{
        ({cos(45)+sin(45)}, {sin(45)}) 
        ({cos(135)+sin(135)}, {sin(135)})
        ({cos(225)+sin(225)}, {sin(225)})
        ({cos(315)+sin(315)}, {sin(315)})};
        \addplot[armygreen, only marks, mark = triangle] coordinates{
        ({cos(45)+2*sin(45)}, {sin(45)}) 
        ({cos(135)+2*sin(135)}, {sin(135)})
        ({cos(225)+2*sin(225)}, {sin(225)})
        ({cos(315)+2*sin(315)}, {sin(315)})};
        %\legend{$a$};
        %\draw[->, thick] (0, -1) -- (-2,-1);
        %\draw[->] (0, 1) -- (2,1);
        \end{axis}
    \end{tikzpicture}
    \caption{A phase-space circle evolved in time by the free Hamiltonian, indicative of the free dispersion of harmonic-oscillator energy eigenstates.}
    \label{fig:sho-free-dispersion}
\end{figure}

The last result shows that the level curves of $W_{\text{free},\, n}(x, p; t)$ at fixed $t$ are ellipses.   As depicted in Figure \ref{fig:sho-free-dispersion}, the curve starts as a circle in $(\tilde x, \tilde p)$ phase space at $t=0$ and evolves according to the free Hamiltonian flow. The evolution is consistent with the dispersion of harmonic-oscillator waveforms in free space, which have widths obeying
\begin{align}
    (\Delta x)^2\, =\, \frac{\hbar}{m\omega} \left(n+\frac{1}{2}\right) \Big[1 + (\omega t)^2\Big]\, ,\quad 
    (\Delta p)^2\, =\, \hbar m\omega \left(n+\frac{1}{2}\right)\ .
\end{align}

%%%%%%%%%%%%%%%%%%%%%

\subsection{Form-preserving transformations of the Moyal equation}

So far we have used the form-preserving transformations of the \Sr\ equation to learn about transforming the corresponding Wigner function. Let us now consider the equation of motion of the Wigner function, the Moyal equation, and attempt to derive its form-preserving transformations directly.  

To write the Moyal equation, we first need a few definitions. The Poisson bracket of 2 phase-space distributions $f(x,p)$ and $g(x,p)$ is written
\begin{align}
\{ f, g \}\ =\ \frac{\del f}{\del x} \frac{\del g}{\del p}\, -\, \frac{\del f}{\del p} \frac{\del g}{\del x}\, =\, f\, \cP\, g\ . 
\end{align}
Here $\cP$ is the Poisson (bi-differential) operator  
\begin{align}
    \cP\ =\ \lpx\,\rpp\, -\, \lpp\,\rpx\ \ ,  
\label{cPdef}\end{align} 
and $f\lpx\,\rpp g$, for example, is defined as $\Pl{x_1}f(x_1,p_1) \,\Pl{x_2} g(x_2,p_2)$ after the identification $(x_1,p_1) = (x_2,p_2) = (x,p)$. 

More generally, a bi-differential operator $\cB$ can be defined by
\begin{align}
    f(x,p)\,\cB\, g(x,p)\ =\ \cB_{12}\, f(x_1, p_1)\, g(x_2, p_2) \left\arrowvert\sub{(x_1, p_2),\, (x_2, p_2) =  (x, p)}\right.\ . 
\label{cPcPij}\end{align}
For example,  
\begin{align}
 \cP_{12}\ =\ \Pl{x_1}\Pl{p_2} \, -\, \Pl{p_1}\Pl{x_2}\ .  
\label{cPij}\end{align}
For short, (\ref{cPcPij}) can be rewritten as \cite{RW18} 
\begin{align}
    f\,\cB\, g\ =\ \cI_{12}\,\cB_{12}\, f_1\, g_2\ ,
\label{onetwo}\end{align}
where $\cI_{12}$ implements the identification $(x_1, p_1) = (x_2, p_2) = (x, p)$.  Note that the identification is done last.

In phase-space quantum mechanics, operator observables become phase-space distributions. 
These so-called symbols are multiplied by a $*$-product homomorphic to the operator product:
\begin{align}
    f\, *\, g\ =\  \cI_{12}\, *_{12}\, f_1\, g_2\ =\  \cI_{12}\, e^{i\hbar\cP_{12}/2}\, f_1\, g_2\ .
\label{star}\end{align}
Incidentally, the notation (\ref{cPcPij}, \ref{onetwo}) can be related to familiar expressions.  With 
\begin{align}
    *_{12}\ =\ e^{i\hbar\cP_{12}/2}\ , 
\end{align}
and substituting 
\begin{align}
    \int dx_1\,dp_1\, \delta(x_1-x)\,\delta(p_1-p) \,  \int dx_2\, dp_2\, \delta(x_2-x)\,\delta(p_2-p)\, 
\end{align}
for the identification $\cI_{12}$ in (\ref{star}), a well-known integral formula for the $*$-product is reproduced (see \cite{ZFC05}, e.g.). 

Operator commutators become $*$-commutators:
\begin{align}
    [ f, g ]_*\, =\, f * g\, -\, g * f\, =\, \cI_{12}\, \left( e^{i\hbar\cP_{12}/2}\, -\, e^{-i\hbar\cP_{12}/2} \right)\, f_1\, g_2\ ,
\end{align}
and the Moyal bracket (or sine bracket) is defined as 
\begin{align}
   \lbb\, f, g\, \rbb\, =\, \frac{[ f, g ]_*}{i\hbar}\, =\, f\, \cM\, g\ ,
\end{align}
so that
\begin{align}
 \cM_{12}\ =\ \frac2\hbar\, \sin\left( \hbar\cP_{12}/2  \right)\ =\ \cP_{12}\,\, {\text{sinc}}\left( \hbar\cP_{12}/2  \right)\ .
\label{cMonetwo}   
\end{align}

If $\cB$ and $\cD$ are 2 bi-differential operators, then   $\cB = \cD$ means 
\begin{align}
    f\,\cB\, g = \cI_{12}\,\cB_{12}\,f_1\,g_2 = \cI_{12}\,\cD_{12}\,f_1\,g_2 = f\,\cD\, g\ ,
\end{align}
for all $f$ and $g$.  Equivalently, $\cB = \cD$ signifies 
\begin{align}
     \cI_{12}\,\cB_{12}\ =\ \cI_{12}\,\cD_{12}\ . 
\end{align} Note, however, that $\cB = \cD$ does {\it not} imply $\cB_{12} = \cD_{12}$. 

We write the Moyal equation, obeyed by the Wigner function $W=W(x,p;t)$, as 
\begin{align}
    \frac{\del W}{\del t}\ =\ \lbb\, H, W\, \rbb\ =\ H\, \cM\, W\ , 
\label{Moyal}\end{align}
and we will consider Hamiltonians of the form 
\begin{align}
    H\, =\, H(x,p;t)\, =\, \frac{p^2}{2m}\, +\, V(x,t)\ .    
\end{align} 

We conjecture that (\ref{WolW}) is the general form-preservation law for Wigner functions.  We know that form-preserving transformations of the \Sr\ equation imply the simple Wigner-function transformation  that shows it behaves as a probability density would when the transformation is canonical. In what follows, we will not assume the transformation of phase-space  coordinates is canonical,  but use only (\ref{WolW}).   

Now consider transforming $(x, p, t) \to (x', p', t')$. The strategy will be to start with the Moyal equation (\ref{Moyal}) for $W=W(x, p; t)$, substitute the relation (\ref{WolW}), to get 
\begin{align}
     \frac{\del {W'}}{\del {t}}\ =\ {H}\, {\cM}\, {W'}\ , 
\label{MoyalWolW}   
\end{align}
and then perform the transformation to the primed variables.  Moyal form-preservation requires that we find the Moyal equation for ${W'} = {W'}({x'}, {p'}; {t'})$, 
\begin{align}
     \frac{\del {W'}}{\del {t'}}\ =\ {H}'\, {\cM}'\, {W'}\ . 
\label{olMoyal}   
\end{align}
We will assume that ${H}' = {p'}^2/2m +{V'}({x'},  {t'})$, and attempt to find form-preserving transformations.    

First, we calculate
\begin{align}
    \frac{p^2}{2m}\,\cM\, W'&\ =\ \frac{1}{2mi\hbar}\,\cI_{12}\, \left( e^{i\hbar\cP_{12}/2}\, -\, e^{-i\hbar\cP_{12}/2} \right)\, p_1^2\, W'_2  \cr
     =&\ \frac{1}{2mi\hbar}\, \cI_{12}\, \bigg[ \Big(p_1 - \frac{i\hbar}2\del_{x_2}\Big)^2 - \Big( p_1 + \frac{i\hbar}2\del_{x_2}\Big)^2 \bigg]\, W'_2 \cr
     =\ & - \frac{p}{m}\,  \del_x W'\ ,
\end{align}
so that (\ref{MoyalWolW}) becomes
\begin{align}
\frac{\del {W'}}{\del {t}}\ =\ - \frac{p}{m}\, \frac{\del{W'}}{\del x}\ +\ {V} \, {\cM}\, {W'}\ . 
\label{MoyalWolWi}       
\end{align}

The Moyal operator $\cM$ of (\ref{MoyalWolWi}) must somehow give rise to  ${\cM}'$. Recall that the form-preserving transformations of the \Sr\ equation (see (\ref{form}), e.g.) exist for arbitrary original potentials. We therefore expect the relation between $\cM$ an $\cM'$ to be independent of the Hamiltonian $H$.  We conjecture that $\cM$ can simply be replaced by $\cM'$ in (\ref{MoyalWolWi}). 

That would be justified if Moyal brackets of any phase-space functions were invariant to the transformations we seek. The transformations would then be quantum analogs of classical canonical transformations, which leave Poisson brackets invariant.  

We therefore proceed by imposing 
\begin{align}
    \cI_{12}\,\cM_{12}\ =\ \cI_{12}\,\cM_{12}'\ ,
\label{IM12IMp}\end{align} or  
\begin{align}
   \cI_{12}\, {\cP}_{12}\,\, {\text{sinc}}(\hbar{\cP}_{12}/2)\ =\ \cI_{12}\, {\cP}'_{12}\,\, {\text{sinc}}(\hbar{\cP}'_{12}/2)\ . 
\label{IM12IMpsinc}\end{align}
Analyzing this condition requires some care. 

First, (\ref{IM12IMp}, \ref{IM12IMpsinc}) would follow from 
\begin{align}
    \cP_{12}\ =\ \cP'_{12}\ , 
\label{cP12cPp}\end{align}
but not from the condition for a canonical transformation, 
\begin{align}
    \cI_{12}\, \cP_{12}\ =\ \cI_{12}\, \cP'_{12}\ .  
\label{IcP12IcPp}\end{align}
The condition (\ref{cP12cPp}) is clearly stronger than (\ref{IcP12IcPp}), and it implies that the transformation is linear, as well as canonical. To show this, recall (\ref{defzeta}) and introduce $\Pl{\zeta} = \left(\begin{matrix} \Pl{x} \cr \Pl{p}\end{matrix}\right)$ to write
\begin{align}
    \cP_{12}\ =\ \Pl{\zeta\sub{1}}^{\, T}\, J\, \Pl{\zeta\sub{2}}\ ,\quad  {\text{where}}\ \ J = \left( \begin{matrix}0 & 1\cr -1 & 0 \end{matrix} \right)\, \ . 
\end{align}
By the chain rule 
\begin{align}
    \Pl{\zeta'}\ =\ \left( \begin{matrix} {\partial x}/{\partial x'}  & {\partial p}/{\partial x'} \cr {\partial x}/{\partial p'} & {\partial p}/{\partial p'}  \end{matrix}  \right)\,  \Pl{\zeta}   \ =:\ \left( {\partial\zeta}/{\partial\zeta'}  \right)\,  \Pl{\zeta} \ . 
\end{align}
(\ref{cP12cPp}) then implies 
\begin{align}
     J\ =\   \big( {\partial\zeta\sub{1}}/{\partial\zeta\sub{1}'}  \big)^T\,   J\,\, \big( {\partial\zeta\sub{2}}/{\partial\zeta\sub{2}'}  \big)\ . 
\label{MoneJMtwo}\end{align}
Without the identification $\cI_{12}$, (\ref{MoneJMtwo}) requires a $\zeta_1, \zeta'_1$-dependence to cancel with a $\zeta_2, \zeta'_2$-dependence. That is impossible unless there is no such dependence, i.e. unless $\left( {\partial\zeta}/{\partial\zeta'}\right)$ doesn't depend on $\zeta'$. That means the transformation is linear. 

A concrete example may be helpful here. Let us describe it in dimensionless variables, denoting them $x$ and $p$, to keep things tidy. Consider the nonlinear canonical transformation 
\begin{align}
    x\, =\, p'\, -\, \left( x' \right)^2\, ,\ \ \ p\, =\, -x'\ .
\label{CTex}\end{align}
For it we find
\begin{align}
    \cP_{12}'\, =\, 2(p_1-p_2)\, \Pl{x_1}\Pl{x_2} \ +\ \cP_{12}\ .
\end{align}
Clearly, $\cP_{12}'\not=\cP_{12}$, and this is a consequence of the nonlinearity of the transformation.  However, $\cI_{12}\cP_{12}' = \cI_{12}\cP_{12}$, i.e. (\ref{IcP12IcPp}), is obeyed, confirming that the transformation is indeed canonical. 

One can further show that 
\begin{align}
  *\sub{12}'\ =\ e^{i\hbar\cP_{12}'/2}\, =\, *\sub{12}\,\, e^{i\hbar\left(p_1-p_2\right)\Pl{x_1}\Pl{x_2}}\, \, e^{-\frac{\hbar^2}{4}\left( \Pl{x_1} + \Pl{x_2} \right)\Pl{x_1}\Pl{x_2}} \ ,
\label{stars12}\end{align}
using the Baker-Campbell-Hausdorff formula.  This implies that 
\begin{align}
    \cM_{12}'\ =\ \frac{1}{i\hbar}\, \left( *'\sub{12}\, -\, *'\sub{21}\right)\ \not=\ \frac{1}{i\hbar}\, \left( *\sub{12}\, -\, *\sub{21}\right)\ =\ \cM_{12}\ .
\end{align}
Finally, upon identifying we obtain
\begin{align}
    \cI_{12}\, 
  *\sub{12}'\ =\ \cI_{12}\, *\sub{12}\, e^{-\frac{\hbar^2}{4}\left( \Pl{x_1} + \Pl{x_2} \right)\Pl{x_1}\Pl{x_2}}  
\end{align} 
from (\ref{stars12}), so that 
\begin{align}
    \cI_{12}\,\cM_{12}'\ \not=\ \cI_{12}\, \cM_{12}\ .
\end{align} 

This example demonstrates that a canonical transformation satisfies (\ref{IcP12IcPp}), but not necessarily (\ref{cP12cPp}).  It also shows that the relation between $\cM_{12}$ and $\cM_{12}'$ is difficult to work out, in general, even from that between  $\cP_{12}$ and $\cP_{12}'$, when known. The canonical transformation (\ref{CTex})  is a special, simple case for which it is possible. 

To proceed, we must restrict our general treatment. In the end we will manage to recover (\ref{form}), and confirm the relationship between the potentials $V$ and ${V'}$ given there.

First, we only consider transformations of the following type: 
\begin{align}
    {x'} = {x'}(x,t)\, ,\ {p'} = {p'}(x,p,t)\, ,\ {t'} = {t'}(t)\, . 
\label{gentran}\end{align}
This ansatz is more general than the transformation (\ref{form}), but note that we do not consider any $p$-dependence for  $x'$.

Second, consider the sinc functions in (\ref{IM12IMpsinc}) as series in $\hbar$. When substituted in the Moyal equation for a specific system, the expansion should be expressible in powers of  dimensionless ratios $\hbar/{s}$, where $s$ is a scale of action characteristic of the system. Changing the system parameters changes $s$, so that if we analyze an ensemble of systems with different parameter values together, $\hbar/s$ can be considered a variable. 

For short, we say that we can consider $\hbar$ to be a continuous parameter. This is consistent with the belief that if $\hbar$ were to have a slightly different value, things wouldn't change drastically. 

With Planck's constant treated as a variable, the coefficients of powers of $\hbar$ on the 2 sides of (\ref{IM12IMpsinc}) must be equal.  This implies 
\begin{align}
    \cI_{12}\,\left( {\cP}_{12} \right)^{1+2n}\ =\ \cI_{12}\,\left( {\cP}'_{12} \right)^{1+2n}\ , \qquad n=0,1,2,\ldots\, .  
\label{cPnPpn}\end{align} 
The lowest-order term reproduces (\ref{IcP12IcPp}), so that the transformation $(x, p)\to (x', p')$ in (\ref{gentran}) must be a canonical one. Therefore we require 
\begin{align}
 \frac{\del{x'}}{\del x}\, \frac{\del{p'}}{\del p}\ =\ 1\  .  
\label{PolPcondition}\end{align}
The momenta $p= m\, dx/dt$ and ${p'} = m\, d{x'}/d{t'}$ are related by
\begin{align}
    {p'}\ =\  \frac{dt}{d{t'}}\, \left( \frac{\del{x'}}{\del x}\, p \ +\ m\, \frac{\del{x'}}{\del t}  \right) \ .
\label{polp}\end{align} 
The condition (\ref{PolPcondition}) requires 
\begin{align}
    \left( \frac{\del{x'}}{\del x} \right)^2\, \frac{dt}{d{t'}}\ =\ 1\ . 
\label{PolPconditioni}\end{align}
If ${t'} = {t'}(t)$, then $\del{x'}/\del x$ must be independent of $x$.  That means ${x'}$ is linear in $x$. We have been led back to transformations of the same type as (\ref{CT}), i.e. linear and canonical. 

Consequently, we will find no Moyal-preserving transformations this way that generalize those relevant to the wave function and its \Sr\ equation. We continue, however, in order to confirm the transformation (\ref{form})  of the potential $V\to V'$ in the phase-space formulation. 

Because the transformation $(x, p) \to (x', p')$ is linear as well as canonical, 
(\ref{cPnPpn}) is satisfied for all $n\in \{0,1,2,\ldots\}$. We can replace $\cM$ with $\cM'$ in (\ref{MoyalWolWi}) to obtain 
\begin{align}
\frac{\del {W'}}{\del {t}}\ =\ - \frac{p}{m}\, \frac{\del{W'}}{\del x}\ +\ {V} \, {\cM}'\, {W'}\ . 
\label{MoyalWolWii}       
\end{align}
The ansatz (\ref{gentran}) yields
\begin{align}
    \frac{\del{W'}}{\del t}\ =&\ \frac{\del{x'}}{\del t}\, \frac{\del{W'}}{\del{x'}}\, +\, \frac{\del{p'}}{\del t}\, \frac{\del{W'}}{\del{p'}}\, +\, \left( \frac{\del{x'}}{\del x} \right)^2\, \frac{\del{W'}}{\del{t'}}\ ,\cr 
    \frac{\del{W'}}{\del x}\ =&\ \frac{\del{x'}}{\del x}\, \frac{\del{W'}}{\del{x'}}\, +\, \frac{\del{p'}}{\del x}\, \frac{\del{W'}}{\del{p'}}\ ,
\end{align}
where (\ref{PolPconditioni}) was used. We relate $p$ to $p'$ using 
\begin{align}
    \frac p m\ =\ \frac{\del x}{\del x'}\, \left( \frac{dt'}{dt}\, \frac{{p'}}{m}\, -\,  \frac{\del x'}{\del t}  \right)\ =\ \frac{\del x'}{\del x}\, \frac{{p'}}{m}\, -\, \frac{\del x}{\del x'}\, \frac{\del  x'}{\del t}\ .
\end{align}
Substituting these relations into (\ref{MoyalWolWii}) gives  
\begin{align}
\frac{\del {W'}}{\del{t'}}\ &=\ - \frac{p'}{m}\, \frac{\del{W'}}{\del x'}\ +\ \left(\frac{\del x}{\del x'}\right)^2{V} \, {\cM}'\, {W'}\ \cr 
-&\ \frac{\del W'}{\del p'}\, \left(\frac{\del x}{\del x'}\right)\, \left[ \frac{p'}{m}\, \frac{\del p'}{\del x}\, +\, \frac{\del x}{\del x'}\, \frac{\del p'}{\del t}\, -\,  \left(\frac{\del x}{\del x'}\right)^2 \frac{\del x'}{\del t}\, \frac{\del p'}{\del x} \right]\, .
\label{MoyalWolWiii}       
\end{align}
Specializing to the transformation (\ref{CT}), the coefficient of $\partial W'/\partial x'$ becomes
\begin{align}
    m\gamma^3\ddot\gamma\, x' \ -\ m\gamma^3\frac{d^2}{dt^2}\big( \gamma\beta \big)\ .
\end{align}
Now 
\begin{align}
    (2\, a\,{x'}\, +\, b)\, \frac{\del W'}{\del p'}\ =\ (a\,{x'}^{\, 2}\, +\, b\, {x'}\, +\, c)\,{\cM}'\, {W'}
\end{align}
for $a,b$ and $c$ independent of ${x'}$ and ${p'}$. Using this and rewriting in terms of $x$ yields 
\begin{align}
\frac{\del {W'}}{\del{t'}}\ &=\ - \frac{p'}{m}\, \frac{\del{W'}}{\del x'}\ +\  \gamma^2 \bigg\{{V} \, +\, m\gamma\ddot\gamma\, \frac{x^2}2\,\cr -&\, m\gamma^2\left( 2\dot\gamma\dot\beta + \gamma\ddot\beta \right) x\, -\, m\gamma\beta\frac{d^2}{dt^2}\left(\gamma\beta\right)\, +\, \frac c{\gamma^2} \bigg\}\, {\cM}'\, {W'}
\, .
\label{MoyalWolWiv}       
\end{align}
We thus find
\begin{align}
    \frac{V'}{\gamma^2}\, =\, V + m\gamma\ddot\gamma \frac{x^2}2 - m\gamma^2\left( 2\dot\gamma\dot\beta + \gamma\ddot\beta \right) x - m\gamma\beta\frac{d^2}{dt^2}\left(\gamma\beta\right) + \frac c{\gamma^2}\, .
\label{VVpMoyal}
\end{align}
This result is consistent with (\ref{form}), given that both $\alpha$ and $c$ are arbitrary functions of $t$. 

Generalization of these $D=1$ phase-space considerations to the $D$-dimensional case is straightforward for the transformation (\ref{3DRI}). But when a $t$-dependent rotation $R$ is included, as in (\ref{Dzeta}), then a vector potential must also be included, so that the gauge properties of the Wigner function must be involved.  We forego the consideration of gauged Wigner functions in this work.

\section{Conclusion}

Symmetries of the nonrelativistic \Sr\ equation include Galilean transformations with parameters depending on time, accompanied by a particular kind of time re-parametrization.   For a free particle, with potential $V=0$, the symmetries are maximal and define what is known as the \Sr\ group \cite{Ni72, Duval2024}. For nonzero potentials, the symmetry is reduced \cite{BSW76}, except for the harmonic oscillator \cite{Ni73}. The \Sr\ equations for vanishing and harmonic potentials can be mapped into each other by transformations similar to the \Sr\ symmetry transformations \cite{Ni73, Ta90}. More generally,  a \Sr\ equation with a potential $V$ can be mapped into another one with a potential $V'$. These \Sr\ form-preserving transformations are the subject of this paper. 

Remarkable wave functions can be produced by these  form-preserving transformations.  That is exemplified here in 1-dimensional ($D=1$) space by discussing 3 examples: the Berry-Balazs accelerating Airy beam \cite{BB79}, the Senitzky coherent excited states \cite{Se54}, and the free dispersion of harmonic-oscillator energy eigenstates. The first and third cases are wave functions in free space found by transforming energy eigenstates, in a linear and harmonic potential, respectively. The Senitzky wave functions are described here as the images of a special time-dependent symmetry transformation for harmonic potentials. 

\Sr\ form-preserving transformations were also studied in $D>1$. Confirming previous results \cite{GRo72, TaI91}, we find that they work in a similar way as in $D=1$, when only a scalar potential is present. When time-dependent rotations are incorporated in the transformations, however, form preservation is possible only if a (transforming) vector potential is included \cite{Nikitin2020}. Our treatment here focuses on determining the explicit, general form of the transformation (\ref{Dformpsi}-\ref{VdimD}) as was done for $D=1$ in (\ref{form}) \cite{Fi99}.

Perhaps most significantly, we formulate and investigate quantum-mechanical form-preserving transformations in phase space, for $D=1$. First, we found that the \Sr\ form preservations (\ref{form}) produced the simple relation (\ref{WolW}) between the original and transformed Wigner functions.  It says that in phase space, the quasi-probability Wigner distribution transforms just as a true probability distribution does.  

We conjecture that the relation (\ref{WolW}) is the general condition for form-preserving transformations of Wigner functions.  It explains and generalizes the connection between the special evolution of curves in phase space and form preservation. The evolution is required by (\ref{WolW}) for the level curves of Wigner functions. For the 3 wave function examples mentioned above the level-curve evolution is depicted in Figures 1-3.

We then searched for form-preserving transformations of the Moyal equation (\ref{Moyal}) directly. The Moyal bracket requires careful treatment.  Canonical transformations do not, in general, leave it invariant. Finding the most general transformations that don't change the Moyal bracket is a difficult problem.  However, by restricting somewhat the possible transformations, as in (\ref{gentran}), we were able to complete the treatment. The form-preserving transformations must then be linear as well as canonical.  We show that those relevant to the \Sr\ equation are the most general transformations of this type that are possible for the Moyal equation. 

To close, let us mention a couple of possible connections of quantum form-preserving transformations to current research.  

First, the transformation from a force-free frame to a uniformly accelerating frame is a simple example of a \Sr\ (and Moyal) form-preserving transformation. By the equivalence principle then, the relation between original and transformed wave functions in (\ref{form}) predicts the phase of a wave function in the presence of uniform gravity. The phase has been verified in the famous neutron-interferometry ``COW'' experiments \cite{COW75} (see \cite{GO80} for a popular treatment). Probing nonuniform gravity would be very interesting, but is not practical.  Gravity aside, it may still be of interest to investigate the wave-function phases induced in noninertial frames, as predicted by form-preserving transformations.  Neutron interferometry is now a well-established technique in quantum studies \cite{RauchWerner15}.  Perhaps form-preserving transformation may also be relevant to  similar investigations using cold-atom interferometry \cite{Dobk25}.

Second, it seems surprising that the \Sr\ equation has a symmetry that is a time-dependent Galilean transformation extended by a certain type of time re-parametrization.  Can one understand better the structure of the \Sr\ form-preserving transformations somehow? Is it a clue to the properties of the non-relativistic limit, with or without gravity? The recent review \cite{Duval2024} makes clear that these and similar questions have been investigated for a long time now. Both group-theoretical/algebraic and geometric approaches  (see also \cite{DSS21} and references therein) are being used.

\section*{Acknowledgements}
This research was supported in part by a Discovery Grant (M.A.W.) from the Natural Sciences and Engineering Research Council (NSERC) of Canada (funding Reference No. RGPIN-2022-04225), and a Discovery Grant supplement from Alberta Innovates and NSERC.

\end{document}